\documentclass[prl,twocolumn,showpacs,preprintnumbers,amsmath,amssymb,floatfix]{revtex4}
\usepackage{graphicx}
\usepackage{dcolumn}
\usepackage{bm}

\begin{document}



\title{Quantum State Control via Trap-induced Shape Resonance \\
in Ultracold Atomic Collisions}

\author{Ren\'e Stock}
\email{restock@info.phys.unm.edu}
\affiliation{{Department of Physics and Astronomy, University of New Mexico, Albuquerque, NM 87131}}
\author{Eric L. Bolda}
\affiliation{{Atomic Physics Division, NIST, Gaithersburg, Maryland 20899-8423}}
\author{Ivan H. Deutsch}
\affiliation{{Department of Physics and Astronomy, University of New Mexico, Albuquerque, NM 87131}}

\date{\today}


\begin{abstract}
We investigate controlled collisions between trapped but separated ultracold
atoms. The interaction between atoms is treated self-consistently using an
energy-dependent delta-function pseudopotential model, whose validity we
establish. At a critical separation, a ``trap-induced shape resonance'' between
a molecular bound states and a vibrational eigenstate of the trap can occur.
This resonance leads to an avoided crossing in the eigenspectrum as a function
of separation. We investigate how this new resonance can be employed for
quantum control.
\end{abstract}


\maketitle


The ability to arbitrarily manipulate the quantum state of a many-body ensemble
represents the ultimate control of a physical system. This task has steadily
advanced in atomic-molecular-optical systems with tremendous progress in
cooling and trapping technology. This has led to the creation of Bose-Einstein
condensates (BEC) and Fermi degenerate gases, and the explorations of new forms
of matter and mesoscopic quantum states previously accessible only in condensed
matter systems \cite{nature:insight}. The addition of engineered traps, such as
optical lattices \cite{Deutsch:98} and other optical \cite{Schlosser:Dumke} and
magnetic \cite{Folman:02} microtraps provides a new knob with which to control
the quantum state. A dramatic example of many-body control in lattices was
demonstrated through the observation of a superfluid to Mott insulator quantum
phase transition and the collapse and revival of the mean field coherence
\cite{Greiner:all}.

The standard approach to modelling and designing coherent states of matter,
such as occur in a quantum phase transition, has its foundations in condensed
matter theory, where one considers solutions to the entire many-body
Hamiltonian. An alternative viewpoint arises from a fundamental theorem of
quantum information theory \cite{Nielsen:Chuang}: an {\em arbitrary} state of a
many-body system can be reached entirely through operations on single bodies
and pairwise interactions. This provides a direct approach to engineering
mesoscopic states through the application of a ``quantum circuit"
\cite{Jane:Duan}. Moreover, one requires only a {\em single} two-body
interaction (e.g. CPHASE or CNOT gate) that entangles the ``particles'' to
contribute to a universal set of quantum logic gates.

In the context of ultracold neutral atoms, whereas manipulating the quantum
state of an individual atom is a very mature technique, arbitrary unitary
mapping of a two-atom system has not yet been achieved. Neutrals, by their very
nature, do not strongly couple to anything. This may be an advantage for
avoiding noise, but it implies that the two-body interaction will generally
require close overlap of the atomic wavepackets. By bringing two atoms within
the same well of a tightly confining microtrap, one can achieve this strong
coupling while remaining in the electronic ground state. Proposals for two-atom
control in such a geometry have been considered using ground state $s$-wave
collisions \cite{Jaksch:99}, Feshbach resonances \cite{Mies:00} and laser
induced Raman transitions \cite{Jaksch:02}. At such close range, the atoms lose
their individual identities and instead must be described as a molecular dimer
which generally does not respect the atomic symmetries. This constrains the
possible encodings of quantum information such that two-body logic gates can be
performed within a well-defined ``logical basis". This constraint can be
overcome by placing the particles in distinguishable locations where the atomic
quantum numbers are conserved asymptotically.  Under typical conditions such
separated atoms would generally encounter very weak interactions.  The coupling
between atoms can be dramatically increased, however, when a resonance of the
two-body system is excited, resulting in long-range interactions. An example of
this is induced electric dipole-dipole interactions associated with excited
electronic states \cite{Brennen:02,Jaksch:00}.

In this letter we describe the physics at the foundation of these protocols by
considering ultracold collisions between trapped but {\em separated} atoms. Our
studies show that the trapping potential can lead to new resonances not found
in free space. Unlike standard atomic collisions, here the relative motion of
the atoms is quantized by the trap, making its description intrinsic to the
process. Resonances can then occur between eigenstates of the trap and
molecular bound states, allowing us to overcome the generally very weak
interactions associated with the van der Waals potential. These ``trap induced
shape resonances'' (TISR) can be substantial but are not accounted for within
perturbation theory, that was applied in previous proposals for quantum logic
via cold collisions \cite{Jaksch:99}.  The TISR provides a new tool for
molecular dimer control (e.g., the production of cold molecules)  and the design
of two-atom quantum logic gates.

\begin{figure}[t]
\includegraphics[width=85mm]{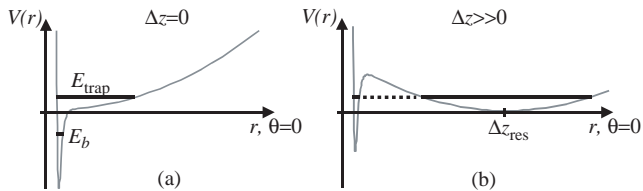}
\caption{Sum of the harmonic trapping potential and chemical binding potential
(gray line) in the relative coordinate $r$ for zero trap separation (a) and
larger trap separation $\Delta z \gg 0$ (b). The molecular bound state at
$E_b$ and trap eigenstate at $E_\mathrm{trap}$ can become resonant at a
critical separation $\Delta z_\mathrm{res}$.}\label{figure1}
\end{figure}

Our model system consists of two trapped and separated atoms that interact
through the molecular potential $V_{\mathrm{int}}$, described by a Hamiltonian,
\begin{eqnarray}
{\hat H} &=&  \frac{ \mathbf{\hat p}_1^2}{2 m}
+ V_{\mathrm{t}}\left({\mathbf{r}_1 + \frac{\Delta \mathbf{z}}{2}}\right)
+ \frac{\mathbf{\hat p}_2^2}{2 m} + V_{\mathrm{t}}\left({\mathbf{r}_2 -\frac{\Delta \mathbf{z}}{2}}\right) \nonumber\\
&&+V_{\mathrm{int}}(\mathbf{r}_1-\mathbf{r}_2)\,,\end{eqnarray}
where $\Delta z$ is the separation of the traps (chosen in the $z$-direction).
The trapping potential $V_{t}$ for the two atoms could be, for example, the
state-dependent trap of a three-dimensional optical lattice potential. In this 
system $\Delta z$ can be continously controlled by the angle between the 
polarization vectors of the counter propagating laser beams 
\cite{Deutsch:98,Mandel:03:2}. We assume atoms are well-localized near a 
potential minimum that is approximated as isotropic and harmonic with 
frequency $\omega$. The two-atom Hamiltonian then separates into one for the
center-of-mass moving in an isotropic harmonic potential, and one for relative 
coordinate dynamics governed by
\begin{equation}\label{Hrel}
{\hat H_{\mathrm{rel}}} = \frac{ \mathbf{\hat p_{\mathrm{rel}}}^2}{2 \mu} +
\frac{1}{2} \mu \omega^2 \left|{\mathbf{r} - \Delta \mathbf{z}}\right|^2 +
V_{\mathrm{int}}(\mathbf{r})\,.
\end{equation}
The reduced mass $\mu = m/2$ moves under the combined effects of a harmonic
trap centered at $\Delta \bf{z}$ and a central interatomic potential
(Fig.~\ref{figure1}).

The harmonic trap is characterized by $z_0 = \left(\hbar/\mu \omega
\right)^{1/2}$, the width of the trap ground wave function, while the
interatomic potential has a much shorter range $x_0$; for ground state van der
Waals interaction $V_{\mathrm{int}}(r)  = -C_6/r^6$ at large interatomic
separation, $x_0 = (2\mu C_6/\hbar^2)^{1/4}/2$ ~\cite{Gribakin:93}. Since for
alkali atoms in typical optical lattices $x_0 \ll z_0$, distortion of the
interatomic potential due to the harmonic trap can be neglected. Furthermore,
for the ultra-cold collisions under consideration here, $s$-wave scattering
dominates the collision. Under these conditions, the interatomic interaction
$V_{\mathrm{int}}(\mathbf{r})$ can be replaced by a zero-range
effective-scattering length pseudopotential
\cite{Huang:57,Bolda:all,Blume:Block}
\begin{equation}
{\hat V}_{\mathrm{eff}}(\mathbf{r}, E_K) = \frac{2 \pi \hbar^2}{\mu}
a_{\mathrm{ eff}}(E_K) \delta(\mathbf r) \frac{\partial}{\partial r}r \,.
\end{equation}
Here $E_K=(\hbar k)^2/(2\mu)$ is the kinetic energy of relative motion for two
atoms in an asymptotic scattering state with momentum $\hbar k$. It determines
an ``energy-dependent scattering length,''
\begin{equation}
 a_{\mathrm{eff}}(E_K) = -\frac{\tan \delta_0(E_K)}{k} \, ,
\label{aeff}
\end{equation}
where $\delta_0(E_K)$ is the $s$-wave collisional phase shift. However, unlike
in a traditional collision where the two atoms are asymptotically free, in this
system the atoms are trapped. The eigenvalues of the system must thus be solved
{\em self-consistently} \cite{Bolda:all}. To this end, the eigenspectrum of the
system is first calculated as a function of the (energy-independent) scattering
length. Second, the effective-scattering length is calculated as a function of
kinetic energy $E_K$ for two {\em untrapped} atoms. The self-consistent energy
eigenvalues are then found numerically as the simultaneous solutions of these
functions. This method has been shown to yield the correct scattering behavior
and energy spectrum for two atoms in a tight harmonic trap \cite{Bolda:all} for
$z_0 \gg x_0$.

Full characterization of the interatomic potential by a pseudopotential
requires not only accurately reproducing the scattering behavior but also the
molecular bound state spectrum. This is particularly important for the TISR
described below. We accomplish this by analytic continuation of the effective
scattering length Eq.~(\ref{aeff}) to {\em{negative}} kinetic energies $E_K =
-\hbar^2 \kappa^2/(2 \mu)$ according to $ a_{\mathrm{eff}}(E_K) = -{\tanh
\imath \delta_0(E_K)}/{\kappa} $, with real and positive $\kappa$. The
self-consistent solutions for these negative energies then accurately reproduce
the entire $s$-wave bound state spectrum as can be easily understood as
follows. Suppose there is a bound state of the actual potential $V_{int}$ at
$E_b = -(\hbar \kappa_b)^2/ (2 \mu)$. The corresponding pole of the $S$-matrix
implies that $\tanh \imath \delta_0(E_b)\rightarrow 1$, and hence
$a_{\mathrm{eff}}(E_b) = 1/\kappa_b$. Now, a pseudopotential with scattering
length $a>0$ possesses exactly one bound state at $E_\delta = -(\hbar^2 /(2 \mu
a^2)$.  By setting $a = a_{\mathrm{eff}}(E_b)$ we recover the exact bound
state, $E_\delta = E_b$. Note that without a self-consistent solution one
cannot choose a pseudopotential  which simultaneously matches the scattering
length and last bound state of the true potential. Moreover, although the
pseudopotential can have no more that one bound state, this self-consistent
solution via the energy-dependent scattering length captures exactly {\em all}
of the $s$-wave bound states, since they are obtained from
$a_{\mathrm{eff}}(E_K)$ and therefore the $S$-matrix.

\begin{figure}[t]
\includegraphics[width=85mm]{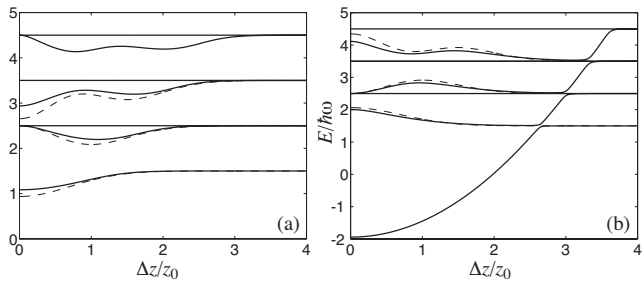}
\caption{Energy spectra as a function of separation $\Delta z$ between traps
for a negative (a) or positive (b) fixed scattering length $a$. The results of
perturbation theory are shown as dashed lines. For $a>0$ one sees the parabolic
energy shift of the molecular bound state due to the harmonic trapping
potential and the avoided crossings associated with the TISR.} \label{figure2}
\end{figure}

With this model we calculate the energy spectrum of Eq. (\ref{Hrel}) for two
trapped but separated atoms. First consider the spectrum for a given scattering
length. We represent the Hamiltonian for arbitrary $\Delta z$ in the basis
corresponding to the solutions with $\Delta z =0$ and a {\em fixed} scattering
length $a$ (i.e. not the self-consistent solution). This basis, derived by
Busch {\em et al.} \cite{Busch:98}, consists of 3D-harmonic-oscillator-like
solutions. The pseudopotential couples only $s$-waves, resulting in irregular
$l=0$ radial waves with singularities at the origin. These solutions also
include the pseudopotential bound state at negative energy. The $l\geq 1$ wave
functions are the regular 3D-harmonic oscillator wave functions. The trap
potential, proportional to $\left| \mathbf{r}-\Delta \mathbf{z} \right|^2=
r^2 - \Delta z \, r \cos\theta + \Delta z^2$ is axially symmetric
and dipolar, thereby preserving the magnetic quantum number of relative motion
and coupling the partial waves $l$ to $l\pm1$. The Hamiltonian is diagonalized
in this basis at each trap separation $\Delta z$. The resulting energy spectrum
for two atoms is shown in Fig.~\ref{figure2} for both positive and negative
scattering lengths, $a=+0.5 z_0$ and $a=-0.5 z_0$, as a function of $\Delta z$.
The results of first-order perturbation theory are also shown for comparison.
For $\Delta z \gg z_0$, we recover the expected unperturbed 3D harmonic
oscillator eigenenergies. As the separation between traps is decreased,
perturbation theory predicts a negative ($a<0$) or positive ($a>0$) energy
shift to the ground state. The expected behavior is seen for $a=-0.5 z_0$, but
an unexpected solution is seen for $a=+0.5 z_0$.

The results for the positive scattering length are explained as follows. For
large positive $a$ there is a molecular bound state close to dissociation. As
$\Delta z$ is increased, the interatomic potential, located at very small
internuclear distances, is pushed up in energy by $\mu \omega^2 \Delta z^2/2$
due to the parabolic trapping potential in Eq.~(\ref{Hrel}). That is, in order
for the separated atoms to collide, they must overcome the potential barrier
created by the trap (see Fig.~\ref{figure1}). When the molecular bound state
becomes resonant with the lowest trap eigenstate, an avoided crossing occurs in
the energy spectrum(see Fig.~\ref{figure2}). As the separation is increased
even further, the molecular bound state becomes resonant with higher-lying trap
states and more  avoided crossings occur. This is a new ``shape resonance'' for
$s$-wave collisions in which the trap barrier plays the role of the centrifugal
barrier in a standard free-space shape resonance for higher partial waves.
Analogous, Feshbach-like, ``confinement induced resonances'' have recently
been  predicted for 1D and 2D trapped bose gases  and for delocalized states
in 3D-optical lattices \cite{Bolda:all,CIR:all}.

\begin{figure}[t]
\includegraphics[width=85mm]{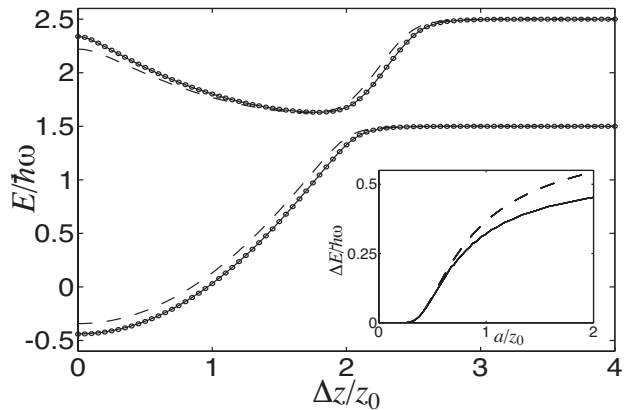}
\caption{Comparison between the energy spectrum of the test step-potential and
that of the pseudopotential approximation. The two lowest energy curves are
shown for the step-potential (solid), the pseudopotential with an energy
dependent scattering length $a_\mathrm{eff}$ (circles) and constant $a$
(dashed). The inset shows the energy gap $\Delta E$ for these two lowest levels
at resonance as a function of scattering length $a$ calculated from energy
spectra at the different scattering lengths. A variational estimate of the
energy gap is shown as the dashed line.} \label{figure3}
\end{figure}

The separation at which the lowest resonance occurs, $\Delta z_\mathrm{res}$,
is easily estimated by equating the sum of the molecular binding energy and
trapping potential at the origin, $E_b +\mu \omega^2 \Delta z^2/2$, to the
vibrational ground state energy of the oscillator, $3\hbar\omega/2$, yielding $
\Delta z_\mathrm{res}/{z_0}=\sqrt{3+{z_0^2}/{a^2}}$. The location and gap of
the avoided crossing depends strongly on the molecular binding energy. For a
deeply bound state, but still positive scattering length, corresponding to $0<a
\ll z_0$, the resonance occurs at much larger separations and with an
exponentially small energy gap. This corresponds to the small probability for
the atoms to tunnel from the trap into the chemical binding potential. Using a
standard variational approach \cite{Merzbacher} based on symmetric and
antisymmetric combinations of the $\Delta z =0$ bound state \cite{Busch:98} and
the trap ground state, we find that for $0<a < 0.2 z_0$ the gap is smaller than
$10^{-4} \hbar \omega$. For $a \gg z_0$ the energy gap asymptotes to a large
value, $\Delta E_\mathrm{max}=0.5640 \hbar\omega$. The shape resonance can therefore be easily
observed for large positive scattering lengths, where the bound state would be
close to dissociation. This can be achieved in tight traps, where the
scattering length is on the order of $z_0$ and the energy gap approaches a
significant fraction of $\hbar \omega$ (see Fig.~\ref{figure3}). For example,
in an optical lattice of $^{133}Cs$ atoms the large scattering lengths of $280
a_0$ to $2400 a_0$ are comparable to typical trap sizes in an optical lattice,
corresponding to a modest Lamb-Dicke parameter $\eta=k_L z_0 =0.1$. A
substantial TISR will result.

To obtain a more accurate spectrum in the case of trapped alkali atoms, we must
account for the energy-dependence of the scattering length in the
self-consistent model described above. As a test case we consider the simplest
possible interatomic potential -- a step-potential of radius $R$ and depth
$V_0$ with a single $s$-wave bound state. The $s$-wave phase shift is given
explicitly by $\delta_0 \left(E_K,V_0\right)= \arctan\left( k\tan(q R)/q\right)
- k R$, where $q=\sqrt{2 \mu (E_K+V_0)/\hbar^2}$\cite{Merzbacher}. The
energy-dependent scattering length is evaluated using Eq.(\ref{aeff}) as a
function of $E_K$. The relative kinetic energy of the colliding atoms is given
by the total energy eigenvalue $E$ minus the trap potential at the origin.

Figure~\ref{figure3} shows the self-consistent energy spectrum as a function of
well separation $\Delta z$. These approximate eigenvalues are compared with the
exact solution for the step-potential ($V_0 = 36.79 \hbar \omega$ and $R = 0.2
z_0$)  plus harmonic potential, calculated numerically. We accomplish this by
expanding the total Hamiltonian in isotropic 3D-harmonic oscillator
wavefunctions and diagonalizing the matrix. Figure~\ref{figure3} also shows a
plot of the constant scattering length approximation, using the zero-energy
scattering length  $a = a_{\mathrm{ eff}}(0)$. As expected, this approximation
fails to capture the correct bound state energy and therefore the correct
location of the shape resonance. In contrast, the self-consistent solution
using the energy-dependent pseudopotential shows excellent agreement with the
exact calculation, even for a well that was chosen to have a fairly long range,
$R=0.2z_0$. The agreement only breaks down when the range of the potential
becomes on the order of trap size, $R> 0.5 z_0$.


Having confirmed the validity of the self-consistent pseudopotential model, we
can use the calculated energy spectrum to design two-atom quantum logic gates.
The use of ultracold collisions was initially considered by Jaksch {\em et
al.}\cite{Jaksch:99} in the non-resonant case, using perturbation theory. Our
analysis shows that in principle, for positive scattering lengths, resonances
will occur at some atomic separation, and perturbation theory will break down.
The resulting avoided crossing in the energy spectrum must be properly
accounted for. This is particularly true for atoms with very large scattering
lengths, such as $^{133}$Cs. The TISR opens the door to new protocols for
entangling two-atom logic gates with separated atoms. For example, a $2\pi$
Rabi oscillation between the trapped atoms and an auxiliary molecular bound
state leads to a phase shift of $-1$ on the two-atom state. If the acquired
phase shift  occurs only for one logical encoding of the atoms, the resulting
unitary transformation is the so called ``CPHASE" two-qubit logic gate
\cite{Nielsen:Chuang}. An optimal regime for operation of this protocol is
where $z_0 \ll a $. In this regime the energy gap (Fig.\ref{figure3} inset)
asymptotes to its maximum value, minimizing the dependence of the cooperative
phase shift on the precise value of the trap width $z_{0}$, and hence reducing errors due to
trap-laser intensity fluctuations.

More generally, beyond quantum logic, the TISR provides a new avenue for
spectroscopy and coherent control of ultracold molecular dimers. Like magnetic
Feshbach resonances, these shape resonances can provide new ultra-high
precision spectroscopic data on the molecular potential \cite{Chin:00}, and the
production of cold molecules tunable by the trap parameters. A complete
characterization of these protocols requires a generalization of our model,
including the full spin-dependent nature of the collision process via the
hyperfine and exchange interactions. We plan to address this in future work
using multichannel Born-Oppenheimer potentials approximated by energy dependent
pseudopotentials.

\begin{acknowledgments}
We thank Paul Julienne, Carl Williams, and Eite Tiesinga for very helpful
discussions. This work was partly supported by the National Security Agency
(NSA) and the Advanced Research and Development Activity (ARDA) under Army
Research Office (ARO) Contract No.~DAAD19-01-1-0648 and by the Office of Naval
Research under Contract No.~N00014-00-1-0575.
\end{acknowledgments}


\bibliographystyle{apsrev}

\end{document}